\documentclass[
 reprint,
superscriptaddress,
rsi,
amsmath,amssymb,
aps,
prx,
]{revtex4-2}

\usepackage{float}
\usepackage[utf8]{inputenc}
\usepackage{mathtools}
\usepackage{verbatim}
\usepackage{physics}
\usepackage{caption}
\usepackage{subcaption}
\usepackage{hyperref}
\usepackage[T1]{fontenc}
\usepackage{color}

\usepackage{hyperref}
\hypersetup{colorlinks=true, citecolor=blue,linkcolor=blue,filecolor=blue,urlcolor=blue}

\begin{document}
\title{ Realistic quantum network simulation for experimental BBM92 key distribution}

\author{Michelle Chalupnik} \email{mchalupnik@alirotech.com}
\affiliation{Aliro Technologies, Inc.,\\ \textnormal{Brighton, MA 02135 USA}}
\author{Brian Doolittle}
\affiliation{Aliro Technologies, Inc.,\\ \textnormal{Brighton, MA 02135 USA}}
\author{Suparna Seshadri}
\affiliation{Aliro Technologies, Inc.,\\ \textnormal{Brighton, MA 02135 USA}}
\author{Eric G. Brown}
\affiliation{Aliro Technologies, Inc.,\\ \textnormal{Brighton, MA 02135 USA}}
\author{Keith Kenemer}
\affiliation{Aliro Technologies, Inc.,\\ \textnormal{Brighton, MA 02135 USA}}
\author{Daniel Winton}
\affiliation{Aliro Technologies, Inc.,\\ \textnormal{Brighton, MA 02135 USA}}
\author{Daniel Sanchez-Rosales}
\affiliation{Aliro Technologies, Inc.,\\ \textnormal{Brighton, MA 02135 USA}}
\author{Matthew Skrzypczyk}
\affiliation{Aliro Technologies, Inc.,\\ \textnormal{Brighton, MA 02135 USA}}
\author{Cara Alexander}
\affiliation{Aliro Technologies, Inc.,\\ \textnormal{Brighton, MA 02135 USA}}
\author{Eric Ostby}
\affiliation{Aliro Technologies, Inc.,\\ \textnormal{Brighton, MA 02135 USA}}
\author{Michael Cubeddu}
\affiliation{Aliro Technologies, Inc.,\\ \textnormal{Brighton, MA 02135 USA}}
\date{March 7, 2026}

\begin{abstract}
Quantum key distribution (QKD) can provide secure key material between two parties without relying on assumptions about the computational power of an eavesdropper. QKD is performed over quantum links and quantum networks, systems which are resource-intensive to deploy and maintain. To evaluate and optimize performance prior to, during, and after deployment, accurate simulations  with attention to physical realism are necessary. Quantum network simulators can simulate a variety of quantum and classical protocols and can assist in quantum network design and optimization by offering realism and flexibility beyond mathematical models which rely on simplifying assumptions and can be intractable to solve as network complexity increases.  We use a versatile discrete  event quantum network simulator to simulate the entanglement-based QKD protocol BBM92 and compare it to our experimental implementation and to existing theory. We find the discrete event quantum network simulator can match experimental key rates and error rates with a lower mean squared error than analytical theory. Furthermore, we simulate secure key rates in a repeater key distribution scenario for which no experimental implementations  exist and find agreement between simulation and analytical theory. Hence, we demonstrate discrete event simulators can meet needs in quantum network simulations which cannot be filled solely by experiment or theory: discrete event simulators can accurately simulate QKD protocols and match experiments in regimes where theoretical models may require more simplifying assumptions, and they can match theoretical models in the opposite scenario where experiments have not yet been performed but theoretical models exist. 
\end{abstract}

\maketitle

\section{Introduction}
\label{sec:intro}

Quantum key distribution (QKD) is a technique for creating a  provably secure secret key between two parties by leveraging properties of quantum mechanics (such as monogamy of entanglement and the no-cloning theorem) \cite{bb84, QKDreviewGisin, E91, decoystates, twinfield}. When used to create a one-time pad, QKD is information-theoretically secure, meaning it does not make any assumptions about the computational resources of an eavesdropper, and the maximum information the eavesdropper can possess can be determined. In contrast, commonly used classical public key cryptography systems like Rivest-Shamir-Adleman (RSA) \cite{RSA}, elliptic curve cryptography (ECC) \cite{koblitz1987elliptic} and even post-quantum cryptography (PQC) \cite{PQC} offer only computational security. Their security is contingent upon the assumption that no efficient algorithms exist to invert the encryption using publicly shared information.

For RSA, an efficient algorithm to factor large numbers can break the encryption and render the encryption method insecure \cite{Shor1994}. An eavesdropper employing the harvest now decrypt later (HNDL) attack can intercept a message sent using RSA encryption and copy and store the encrypted message and public key until a future date when quantum computers have advanced enough to allow efficient decryption. Messages encrypted using PQC are similarly vulnerable. Although no efficient algorithms have yet been discovered to break the lattice-based cryptography underpinning PQC, HNDL attacks can still be levied on PQC systems in the same manner \cite{SIDHAttack}. 

As both classical and quantum computers continue to increase in size and connectivity, quantum-enabled encryption can  exceed the security offered by classical encryption methods by providing security that will last in perpetuity. Because of the no-cloning property of quantum mechanics and the absence of exploitable mathematical structure in the resulting key material, messages encrypted with quantum keys are future-proof, as they cannot be copied and decrypted at a later date even given future advancements in computing technologies.

A variety of QKD protocols have been demonstrated over long distances \cite{ ursin2007, yin2020, wang2022, Liu1000km} with fiber multiplexing \cite{multiplex2018, Beutel2022}, via satellite \cite{ecker2021, satellite2025}  and with secure authentication and integrated optical switches \cite{yongswitches2021}. When integrated in a quantum network with quantum memories, quantum error correction can be performed prior to measurements \cite{Feihureview2020}. However, QKD protocols  vary in their security assumptions and vulnerability to attacks. 

Bennett-Brassard-Mermin 1992 (BBM92) is an entanglement-based QKD protocol whereby Alice and Bob derive each bit in a shared secret key by measuring each  photon from an entangled Bell pair in a basis randomly selected from two orthogonal bases \cite{BBM92}. BBM92 lacks vulnerabilities present in other QKD schemes because it uses an entangled photon source to distribute the same maximally entangled Bell state to produce each key bit. In particular, QKD security assumptions including that the encoded state emitted from the source is  basis-independent can be verified for the BBM92 protocol even if the source is  controlled by an eavesdropper  \cite{Feihureview2020}. High entanglement fidelity is a key requirement behind BBM92 security, and in deployed BBM92, maintaining a low quantum bit error rate (QBER) ensures the security of the protocol. 

Although analytical models exist and can provide useful estimates for secure key rates and other important metrics for BBM92 in simple quantum networks, the theory  becomes intractable as the system complexity increases. In contrast, quantum network simulators can simulate parameter regimes which are not feasible to achieve with current-day devices, provide insights during experimental troubleshooting, and save the time and cost of a full physical experimental implementation. Particularly as quantum networks move towards full stack implementations, quantum network simulators can assist in simulation and optimization as quantum network protocols grow in complexity.  

In this paper, we accurately simulate BBM92 key exchange using the Aliro Quantum Network Simulator (AQNSim) \cite{alirosimulatorpub}, a  versatile and adaptable discrete  event  Python simulation package for quantum networks. We choose to use AQNSim for our simulations because of its high degree of customizability, physical accuracy, detailed component modeling, and flexible abstraction. While we do not perform a systematic comparison across all available simulators, we use AQNSim as a reference implementation to show that a Python-based discrete event quantum network simulation framework can be used to accurately and realistically simulate quantum network protocols. We compare our simulation results to our  experimental implementation of BBM92 and to an existing numerical model for continuous-wave-pumped BBM92 \cite{NeumannPRA}. We find that simulated results using AQNSim match our experiments as well as the theoretical model in regimes where the theoretical model is valid. In regimes where assumptions required for the theoretical model fail, our simulations match experimental data with a lower average mean squared error. Moreover, we leverage AQNSim to simulate entanglement generation rates and secure key distribution rates in a repeater network with multiple links, showing the utility of AQNSim for modeling scenarios where experiments may still be unfeasible and where analytical models may only exist under fixed assumptions.

\section{BBM92 Experimental Setup}
\label{sec:exp}

\begin{figure}[tb]
\includegraphics[width=8.0cm]{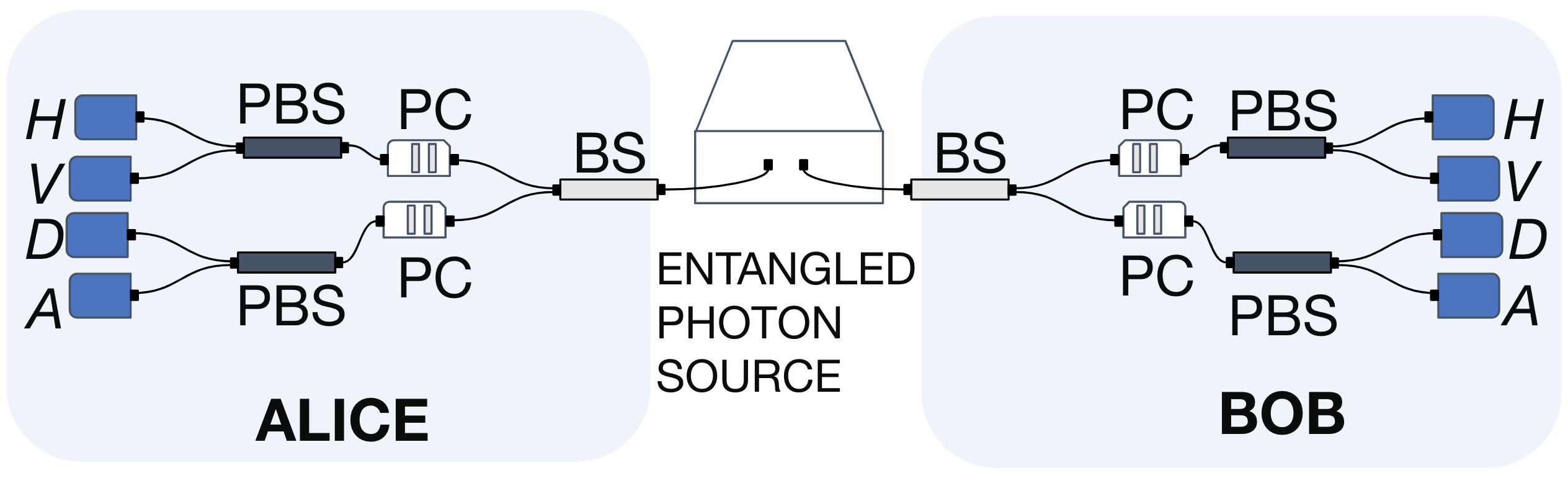}
\caption{Schematic of the BBM92 experimental setup used to produce keys, and evaluate key rates and QBER metrics. Detectors are shown in dark blue and are labeled by polarization measurement outcome. BS: beam splitter, PC: polarization control and correction, PBS: polarizing beam splitter. 
\label{fig:setup}}
\end{figure}

We implemented BBM92 key exchange using continuous-wave pumped, near-infrared polarization-entangled photons. In our experimental setup, Alice and Bob generate a shared secret key by measuring photons emitted from an entangled photon source (see Figure \ref{fig:setup}). A narrowband entangled photon source emits polarization-entangled photons in the quantum state


\begin{align}
\label{eq:quantumstate}
\ket{\Psi^+} = \frac{1}{\sqrt{2}} \left( \ket{HV} + \ket{VH}\right),
\end{align}
where  $\ket{H}$ and $\ket{V}$ form an orthonormal basis of horizontal and vertical single photon polarization states.
The commercially-available source contains a laser which pumps two type-II spontaneous parametric down conversion (SPDC) quasi-phase-matched crystals in a Mach-Zehnder interferometer configuration (see Appendix A). The source emits photons at nominally 810 nm with a full-width half maximum (FWHM) of approximately 3 nm.

The entangled photons travel in separate single-mode fibers in a laboratory to each communication partner, Alice and Bob. At each communication partner,  a fiber-based 50:50 beam splitter acts as a passive switch between an $HV$ basis measurement and an $AD$ basis measurement, where 
$\ket{D}$$ = \frac{1}{\sqrt{2}}\left( \ket{H} + \ket{V}\right)$ and $\ket{A} = \frac{1}{\sqrt{2}}\left( \ket{H} - \ket{V} \right)$.  Polarization compensation is performed to correct for polarization drift (see the measurement correlation matrix in Appendix B). At each communication partner, in-fiber polarization beam splitters route the received photon to one of four single-photon avalanche detectors for measurement. A time tagger with root-mean-square jitter of nominally 42 ps is used to collect and timetag detection events. Synchronization between Alice's detectors and Bob's detectors is necessary for BBM92 key exchange and is performed by characterizing the relative time delays for photons traveling to each detector (see Appendix A).

To perform key exchange, time-tagged detection events are collected at each detector.   Alice and Bob publicly exchange detection timestamps (adjusted for delays between detectors) labeled only by the randomly selected measurement basis. Alice and Bob then keep only the timestamped detection results which they both have measured in the same basis  ("sifting"). A shared raw key between Alice and Bob is then distilled according to the detection results and anticipated correlations given the distributed Bell state (equation \ref{eq:quantumstate}). The publicly exchanged basis information does not reveal any information about the secret key which would compromise the security, and any modification of the Bell state which is sent to Alice and Bob will result in a detectable increase in QBER. In this work, the QBER is calculated by comparing Alice and Bob's entire raw keys. In a production-ready deployment, a fraction of the key is sacrificed to estimate the QBER of the entire key. Following the estimation of the QBER, error correction can be performed (using, for instance, low density parity check codes \cite{LDPC}) followed by error verification and  authentication. As a final step, privacy amplification can minimize the information extractable by an eavesdropper \cite{Feihureview2020}.

\section{Discrete Event Simulation for  Quantum Networks}
\label{sec:aqnsim}

Discrete event simulation of quantum networks combines a full quantum mechanical  treatment of qubits with event-based logic and timing for quantum and classical signals. Discrete event quantum network simulators such as SeQUeNCe \cite{sequencepaper}, QuISP \cite{quisp}, NetSquid \cite{netsquid}, and others \cite{QNSimulatorsReview} have accelerated quantum network protocol design. For existing discrete event quantum network simulators, quantum memories, qubits, and quantum gates are often highly abstracted from their physical representations or else specific to a given physical implementation. In comparison, AQNSim is a quantum network discrete event simulation Python package which provides realistic modeling of quantum optics with customizable abstraction. Depending on the experiment and application, this realism can be critical to accurately modeling  qubit behavior in a quantum network. For BBM92, hardware specifications like timing jitter, detector dead times, and dark counts all are necessary for accurate modeling of key rates and QBERs and can be incorporated using AQNSim. 

AQNSim can simulate the physics and timing of quantum key distribution protocols, entanglement swapping, quantum gates and quantum operators, quantum optical components including beam splitters and waveplates, and more at a user-specified level of abstraction. The event-based nature of AQNSim supports the simulation of classical messages and communication on the network. When needed for increased performance, AQNSim also has a multi-shot capability, allowing many measurements to be performed in parallel by calculating the pre-measurement quantum state once and sampling from it many times. 

\begin{figure}[tb]
\includegraphics[width=8.0cm]{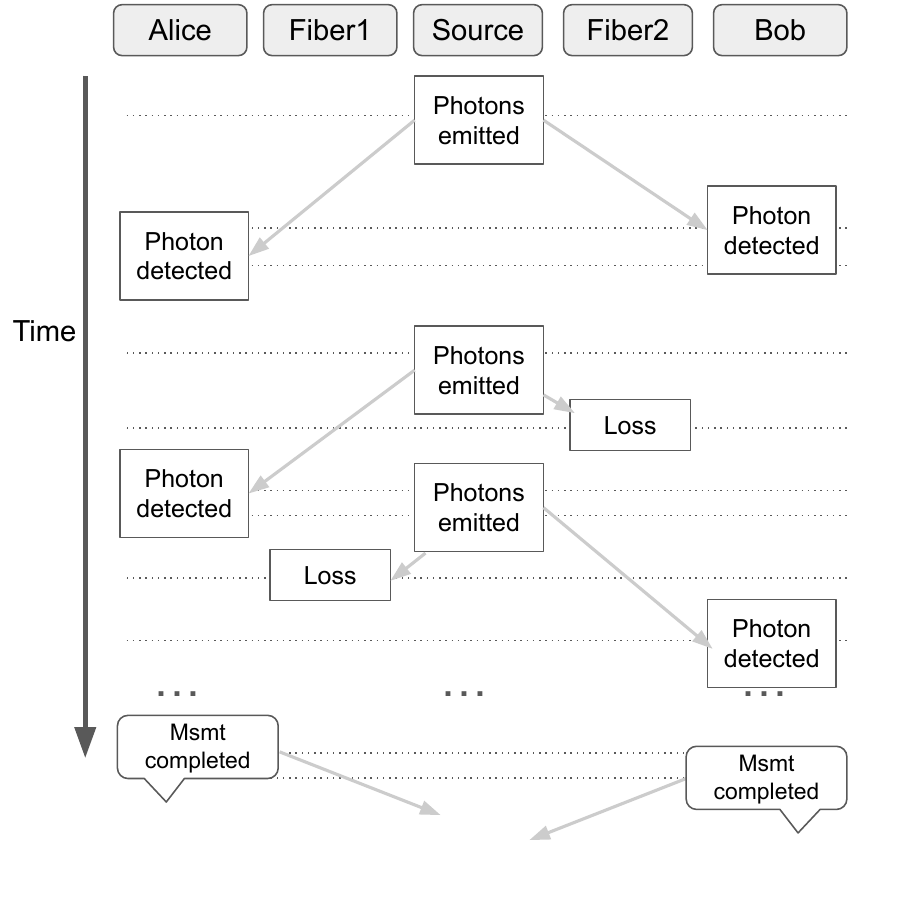}
\caption{Schematic illustrating the discrete event simulation for BBM92 key exchange. An entangled photon pair is emitted from the source. Each photon may be lost during transmission in a loss event or detected by Alice or Bob in a detection event which may be shifted due to detector jitter. Diagonal gray arrows show  signal propagation. \label{fig:aqnsim}} 
\end{figure}

The discrete event framework of AQNSim  enables the accurate and complete simulation of all timestamped events which occur between two communication partners during the BBM92 protocol (see Figure \ref{fig:aqnsim}). To simulate BBM92 key exchange, three nodes (Alice, Bob, and the source node) are created within a network, and classical links and fiber links are created and configured appropriately to allow classical and quantum communications between the three nodes.

An entangled photon source is initialized in the source node and entangled photon pairs are emitted from the entangled photon source as Werner states. A Werner state is
\begin{align} 
W(F)= \frac{4 F - 1}{3} \ket{\Psi^+}\bra{\Psi^+} + \frac{1 - F}{3} I,
\end{align}
where $F$ is the fidelity between the  quantum state measured at Alice and Bob and the desired Bell state. 

 We adopt the Werner state model  because we assume that our generated entangled photon pairs are approximately subject to isotropic  depolarizing noise. Isotropic depolarizing noise replaces the maximally entangled state with the maximally mixed state with some probability and is often used to approximate generalized phenomenological qubit noise without requiring a full determination of the exact noise channel that the qubit experiences  \cite{nielsenChuang}.
 We find this approximation reproduces the experimental data with a high degree of accuracy without requiring full process tomography.

When the probability of measuring an accidental coincidence between photons from different entangled pairs is small compared to the probability of measuring a coincidence between photons from the same entangled pair, $F$ is related to the QBER of the measurement results in each basis ($X$, $Y$, $Z$) by \cite{fidelityqber}
\begin{align} 
F = 1- \frac{\text{QBER}_X + \text{QBER}_Y + \text{QBER}_Z}{2}.
\end{align}
Visibility $V$ relates to the QBER via the equation $V = 1 - 2\;\text{QBER}$. When a Werner state is assumed as the quantum state, the expression for fidelity can be simplified to
\begin{align} 
F = 1- \frac{3}{2}\text{QBER}.
\end{align}
For the simulations, we find $F$ from the measured QBER from our experiments (see Appendix C for further details).


The receiving nodes in the simulation contain the optical equipment at Alice and Bob used to measure photons. For each communication partner, one beam splitter, two polarizing beam splitters, one half waveplate, and four photon detectors are connected to simulate the setup shown in Figure \ref{fig:setup}. Input and output ports on each component allow for routing of photons, and loss can be specified for each component.

Each node has an associated asynchronous protocol that determines the node's actions  while the discrete event simulation runs. Alice and Bob each follow a protocol with steps for receiving and measuring photons, while the entangled photon source operates under a protocol with instructions for sending entangled photons to Alice and Bob. Simulations are accelerated using  the deferred measurements capability of AQNSim's quantum simulation engine. The deferred measurements capability allows the discrete  event simulation to be augmented with parallel simulations of many instances of the same tree of events.

\section{Theoretical Model for BBM92 Key Generation}
\label{sec:theory}

Existing theoretical models can predict raw key rates, secure key rates, and QBERs for continuous-wave-pumped BBM92 \cite{NeumannPRA}. The theoretical model for key generation rates for continuous-wave-pumped BBM92 is a modification of the theoretical model for key generation rates for pulsed-pump BBM92 using parametric down conversion sources \cite{Ma2007PulsedBBM92}. In this section, we follow these models to produce theoretical estimates of key rates and QBERs which will then be compared with our BBM92 experimental results and simulation results. 

Alice and Bob wish to establish a secret key using the BBM92 setup shown in Figure \ref{fig:setup}. A continuous wave entangled photon source stochastically emits polarization-entangled photon pairs (see equation \ref{eq:quantumstate}) at an average rate of $B$ pairs per second, where $B$ is the source brightness. One photon from the pair is routed to Alice and the other photon to Bob. All losses on the link between the entangled photon source and Alice (Bob) other than losses caused by detector efficiencies, dead times, and finite detection resolution are represented by the efficiency $\eta_A$ ($\eta_B$). 

The measured singles count rates at Alice's detectors ($S_A$) and Bob's detectors ($S_B$) are equal to 
\begin{align} 
S_A = B \eta_A \eta_{D} \eta_{dt, A} + D_A \\
S_B = B \eta_B \eta_{D} \eta_{dt, B} + D_B,
\end{align}
where $\eta_{D}$ is the maximum detection efficiency at 810 nm of each detector  and which is approximately the same for each detector, $\eta_{dt, A}$ ($\eta_{dt, B}$) is the contribution to efficiency due to the detector dead time at Alice (Bob) and $D_A$ ($D_B$) are the summed average dark counts per second for each of Alice's (Bob's) detectors. Count rates here are reported in counts per second rather than hertz, to emphasize their stochastic nature. 

The contribution to efficiency from the detector dead time, $\eta_{dt}$, can be calculated as
\begin{align} 
\eta_{dt, i} = \frac{1}{1 + (B \eta_i \eta_{D} + D_i) t_d /n_d},
\end{align}
where $n_d$ is the number of detectors per key exchange participant and $t_d$ is the dead time.  

Photon coincidence rates between Alice and Bob are predicted in the analytical model by considering a theoretical histogram showing differences between arrival times between detections at Alice and detections at Bob. The total number of coincidences in a coincidence window $t_c$ is thus given by the integral of the coincidence histogram from $\mu - t_c/2$ to $\mu + t_c/2$, where $\mu$ is the offset of the histogram peak along the x-axis and $t_c$ is the coincidence window size. Coincidence rates can be measured as this integral divided by total acquisition time and $t_c$. 

At long collection times, the shape of the coincidence histogram will be Gaussian-like with a FWHM equal to the detection resolution time, $t_r$. Detection resolution encompasses the spread in photon arrival time caused by detector jitter, chromatic dispersion, and finite photon coherence time. By approximating the histogram as a Gaussian with a FWHM equal to $t_r$ and integrating the Gaussian over $t_c$, an efficiency factor for coincidences can be obtained as
\begin{align} 
\eta_r = \text{erf}\left[ \sqrt{\ln(2)} \frac{t_c}{t_r} \right].
\end{align}
When $t_r \ll t_c$, detection jitter is negligible and $\eta_r \rightarrow 1$.

The coincidence rate between Alice and Bob ($R_{\text{coin, pairs}}$), not including any accidental coincidences which would result from two distinct entangled pairs happening to fall within the same coincidence window, will thus be given by
\begin{align} 
R_{\text{coin, pairs}} = B \, \eta_A \, \eta_B \, \eta_D^2 \, \eta_{dt, A} \, \eta_{dt, B} \, \eta_r.
\end{align}

The total measured coincidence rate will also include accidental coincidences. These accidental coincidences include cases where two photons from two different entangled pairs fall in the same time bin, or a photon emitted from the entangled source and a dark count photon fall in the same time bin. Half of these accidental coincidences will, by chance, result in valid key bits and not contribute to the QBER. 

The probability of the entangled photon source emitting $n$ photon pairs in a given coincidence window follows a Poisson distribution \cite{QKDreviewGisin}. For a Poisson point process, the probability of $n$ events occurring in a time period given the average rate of event per time period $\lambda$ is 
\begin{align}
P(n, \lambda) = \frac{\lambda^n}{n!}e^ {- \lambda}.
\end{align}
The probability that a detector detects $n$ dark counts in a given coincidence window also can be approximated using Poissonian statistics.

Because the probabilities of each photon pair emission event are independent, a good approximation for the probability that photon detection events from photons belonging to different entangled pairs occur in a given coincidence window is therefore the probability of one or more photon detection events occurring in the coincidence window on Alice's detectors multiplied by the probability of one or more photon detection events occurring in the coincidence window on Bob's detectors, 
\begin{align}
P_{\text{acc}, t_c} \approx \left(1 - e^{-S_A t_c} \right) \left(1 - e^{-S_B t_c} \right).
\end{align}
This expression is valid in the regime where singles count rates are high compared to coincidence count rates, but when the ratio is low, it overestimates the rate of accidental coincidences because it includes coincidences from true pairs.

The rate of accidental coincidences between Alice and Bob is then
\begin{align}
R_{\text{acc}} =  \frac{P_{\text{acc}, t_c}}{t_c}.
\end{align}
and the rate of total coincidences is 
\begin{align}
R_{\text{coin}} = R_{\text{coin, pairs}} + R_{\text{acc}}.
\end{align}

For BBM92 with passive random basis selection using a non-polarizing 50:50 beam splitter, the raw key rate will be one half of the coincidence count rate,
\begin{align} 
R_{\text{key, raw}} = \frac{1}{2} R_{\text{coin}}.
\end{align}
The raw key rate is the rate of key bits produced after basis selection and sifting but before error correction.

To find the probability of error per bit, or QBER, several different potential sources of error must be considered. These sources of error include the error  due to imperfections in the optical setup ($R_{\text{opt, err}}$), and the error  from accidental coincidences that do not produce valid key bits ($R_{\text{acc, err}}$).

Imperfections in the optical setup contribute to error by causing the quantum state which reaches Alice and Bob's photon detectors to deviate from the desired quantum state  (equation \ref{eq:quantumstate}). Defects in measurement optics like waveplates or polarization beam splitters, imperfect polarization drift compensation, fluctuations in temperature, and small misalignments in quasi-phase matching within the integrated source will lower the visibility of the measured quantum state. 

The error rate from imperfections in the optical setup is equal to
\begin{align} 
R_{\text{opt, err}} = \frac{1}{2} R_{\text{coin}} p_o,
\end{align}
where $p_o$ is the probability of bit flip for a given bit in the key due to imperfect optics, and the factor of one half comes from the basis selection and sifting step. $p_o$ will approximately equal the QBER when bit errors from accidental coincidences are negligible. 

The error rate from accidental coincidences will be equal to one quarter of the accidental coincidence rate, with the factor of one quarter accounting for sifting as well as for one half of the remaining accidental coincidences producing valid key bits,
\begin{align} 
R_{\text{acc, err}} = \frac{1}{4} R_{\text{acc}}.
\end{align}

The overall error rate is therefore equal to
\begin{align} 
\text{QBER} = \frac{ R_{\text{opt, err}} + R_{\text{acc, err}}}{R_{\text{key, raw}}}.
\end{align}

Error correction protocols in deployed BBM92 systems will consume valid key bits in order to detect and reconcile key errors between Alice and Bob. With error correction, secure key distribution can still be performed in the presence of errors and a theoretical bound on the maximum secure key rate given an infinitely long key length, the asymptotic secure key rate, can be determined \cite{Shor2000SecurityProof}. For an equal probability of measurement in either the $HV$ or $AD$ basis, and for a conservative assumption on the amount of key consumed to detect and rectify errors which accounts for finite key block sizes in the error correction process, we can predict the asymptotic secure key rate as \cite{Elkouss2011}
\begin{align} 
R_{\text{key, sec}} = R_{\text{key, raw}} \left[ 1 - 2.1 \, H(\text{QBER}) \right],
\label{eq:securekeyrate}
\end{align}
where $H(x)$ is the binary entropy function,
\begin{align} 
H(x) = -x \log_2(x) - (1-x) \log_2 (1-x).
\end{align}
For finite key lengths, the value of the QBER (needed for choosing the most efficient error correction code and for privacy amplification) must be estimated by sampling bits from the key. From the estimated QBER the secure key rate for a finite key can then be probabilistically bounded.

\section{Comparing Simulation, Experiment, and Theory for BBM92 Key Generation}

\begin{figure*}[tb]
\includegraphics[width=17.4cm] {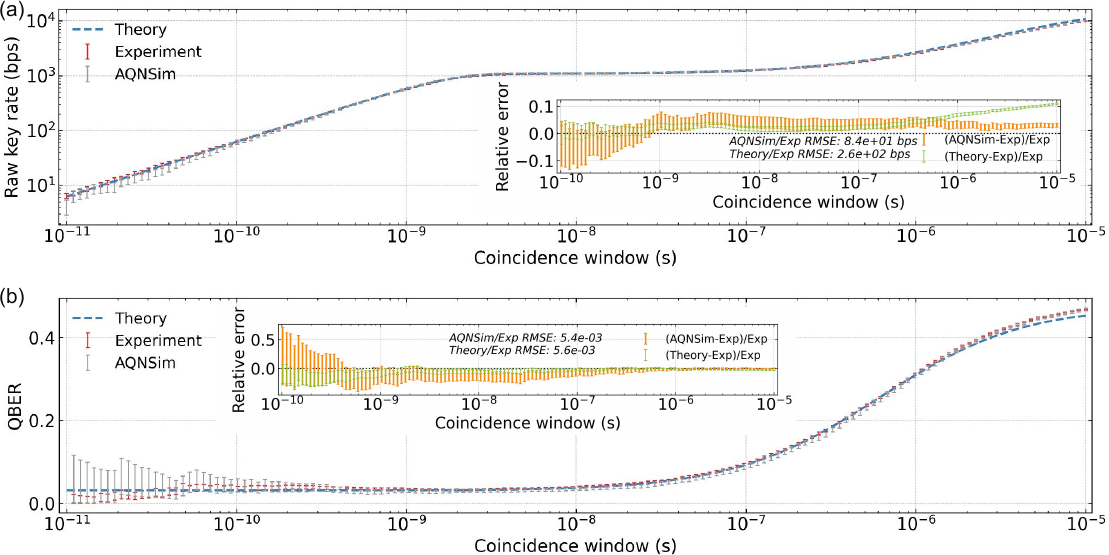}
\caption{ (a) Raw key rates (with relative errors as inset plot) and (b) QBERs (with relative errors as inset plot) for theory, experiment, and AQNSim simulation of BBM92 key exchange. Relevant experimental parameters used in simulations are provided in Table \ref{tab:parameters}. The AQNSim key rates and QBERs are obtained from  a  randomly seeded simulation, re-binned at each coincidence window, averaged over 4,000,000 shots. The experimental key rates and QBERs are obtained from a single experimental run, with the data re-binned at each coincidence window size. Root mean squared errors are displayed for AQNSim and experiment and for theory and experiment for coincidence windows greater than 100 picoseconds. } 
\label{fig:keyratesandqbers}
\end{figure*}

We implement BBM92 key exchange between Alice and Bob in a laboratory setting, and use AQNSim (see section \ref{sec:aqnsim}) and a  theoretical model (see section \ref{sec:theory}) to accurately model the BBM92 quantum key distribution between Alice and Bob. 

\begin{figure*}[tb]
\includegraphics[width=17.0cm]{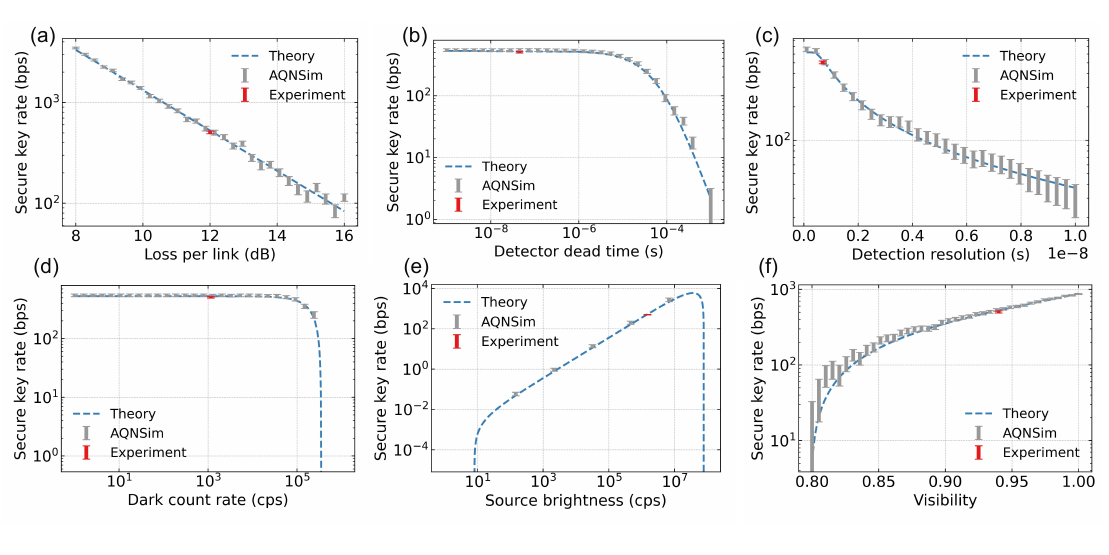}
\caption{ Asymptotic BBM92 secure key rate for simulation, theory, and experiment for swept parameters (a) loss per link, (b) detector dead time, (c) detection resolution, (d) dark count rate per communication partner, (e) source brightness, and (f) visibility. The coincidence window for all plots is 2 nanoseconds and all other parameters are given in Table \ref{tab:parameters}. }
\label{fig:theorysweeps}
\end{figure*}

Raw key rates and QBERs for the experiment for different coincidence window durations are plotted in Figure \ref{fig:keyratesandqbers}.  The experimental data plotted in red is re-binned at each coincidence window  to calculate the raw key rate and QBER. For the AQNSim model and for the analytical model of the BBM92 key exchange,  parameters such as internal source brightness, loss per link, detection resolution, and entangled photon source and optical setup visibility are estimated from the experimental data using methods described in Appendix C. The experimentally measured QBER at 1 nanosecond is used to set the fidelity of the Werner state for the simulated and analytical models and this fidelity is verified with independent entangled photon source visibility measurements. Other relevant parameters, including detector dead time, are obtained from equipment data sheets. These parameters are summarized in Table \ref{tab:parameters}. 

Using the parameters in Table \ref{tab:parameters}, key rates and QBERs are  simulated using an AQNSim replica of the BBM92 setup (see Figure \ref{fig:keyratesandqbers}). To improve simulation speed, entangled pair emission events are simulated in parallel, with a single shot consisting of a single pair emission event, and with the stochastic delays between pair emission events drawn from a Poisson distribution (event rate $\lambda = B$ cps). Additionally, theoretical predictions for raw key rates and QBERs are plotted in Figure \ref{fig:keyratesandqbers}. 

Theory, experiment, and simulation show excellent agreement. Relative errors for the raw key rate and QBER for both the theoretical model as well as for AQNSim simulation are plotted in insets in Figure \ref{fig:keyratesandqbers} for coincidence windows of 100 picoseconds or larger. At smaller coincidence windows the error bars are larger because the number of bits in the simulated key and in the experimentally derived key decreases.  AQNSim simulation predicts the QBER within 1 $\sigma$ error margins and predicts the raw key rates within a few percent. The deviation of simulation from experiment is near the tolerance expected given the uncertainties on the estimated parameters used in simulation. The inset in Figure \ref{fig:keyratesandqbers}a shows the theoretical model displaying a growing deviation for the predicted raw key rate from the experimentally measured raw key rate as coincidence windows increase in size (beginning around coincidence windows greater than 1 microsecond). Deviation of experimental results from theory occurs  from larger coincidence windows leading to more accidental coincidences which eliminate true coincidences by falling in the same coincidence window as well as from an overestimation in accidental coincidences in the theoretical model by counting true coincidences as accidental.

AQNSim can predict BBM92 asymptotic secure key rates for a wide range of equipment  specifications and operating regimes.  We simulate the asymptotic secure key rate for our BBM92 setup using AQNSim and plot the rate against different experimental parameters including source brightness, dark count rate, detector dead time, detection resolution, loss per link, and visibility (see Figure \ref{fig:theorysweeps}). Here, the secure key rate is calculated from the raw key rate, QBER, and equation \ref{eq:securekeyrate}. In Figure \ref{fig:theorysweeps}  the AQNSim-simulated  secure key rates match theoretical predictions and experiment. AQNSim can accurately predict the secure key rate for BBM92 and can assist in finding trends and tolerances for experimental parameters.

\begin{table}[]
\centering
\caption{Summary of relevant BBM92 experimental parameters. \label{tab:parameters}}
\begin{tabular}{|l|l|}
\hline
\textbf{Measured Parameter}                                                                                  & \textbf{Value}     \\ \hline
Source visibility                                                                                 & 94\%              \\ \hline
Source wavelength                                                                                 & 810 nm              \\ \hline
Source bandwidth (FWHM)                                                                                 & 3 nm              \\ \hline
Source pair emission rate                                                                                   & 1.50E6 cps               \\ \hline
Detection resolution (FWHM)                                                                                & 1600 ps              \\ \hline
Detector dead time                                                                              & 45 ns               \\ \hline
Detector maximum efficiency                                                                              & 0.60                \\ \hline
Detector dark counts (Alice)                                                                                 & 500 cps     \\ \hline
Detector dark counts (Bob)                                                                                 & 1800 cps              \\ \hline

Alice link loss                                                                                  & 12.0 dB               \\ \hline
Bob link loss                                                                                   & 12.0 dB               \\ \hline
\end{tabular}
\end{table}

\section{Comparing Simulation and Theory for BBM92 in  Quantum Repeaters}
Photons traveling in a fiber experience loss, and the probability of photon loss increases exponentially with the length of the fiber. This photon loss together with detector dark counts limit the maximum distance over which quantum secure communication can be deployed. To counteract this loss, trusted relay nodes have been proposed for prepare-and-measure QKD protocols like BB84 \cite{elliott2005, trustedRelayNodePRA}. Trusted relay nodes can extend the distance over which prepare-and-measure QKD protocols can be performed, but their security relies on the assumption that an eavesdropper cannot access and manipulate the relay node.

To enable quantum secure communication between two parties over long distances without security assumptions requiring trusted relay nodes, quantum repeaters have been proposed \cite{Briegel1998, repeaterReview, atomicensembles}. In a quantum repeater network, the quantum channel between Alice and Bob is divided into segments (elementary links), with a quantum repeater node placed between each elementary link. Entanglement is first generated across the adjacent elementary links and stored in quantum memories, and then established between Alice and Bob by applying entanglement swapping to each quantum memory in each quantum repeater node. To obtain the shared secret bit, Alice and Bob then each measure the qubit stored in their quantum memory.

\begin{figure}[tb]
\includegraphics[width=8.0cm]{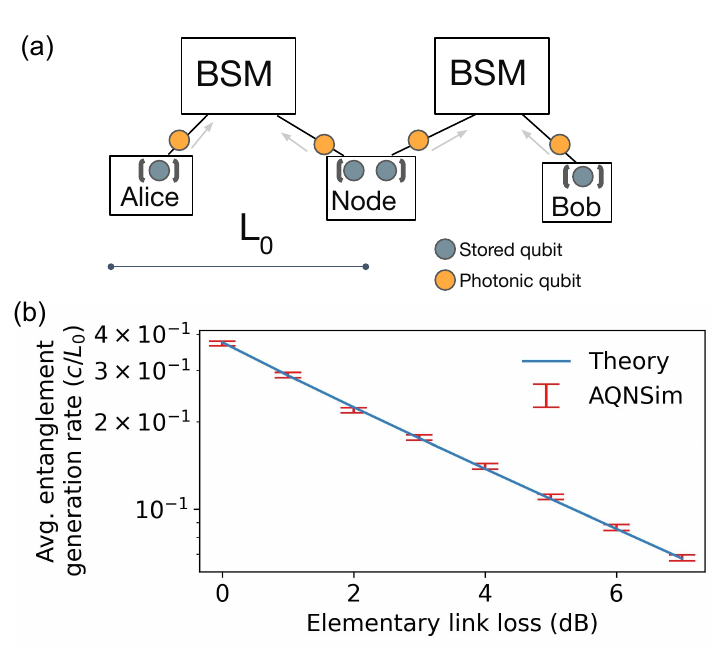}
\caption{(a) Entanglement creation using a repeater containing two end nodes (Alice and Bob), one central repeater node with two stored qubits capable  of two-qubit gate operations ($r=1)$, and two Bell state measurement stations (BSM).  (b) The average entanglement generation rate (in units of $c/L_0$) for different elementary link loss values calculated from simulation (using AQNSim with 1000 shots) and theory for a two-link repeater.  } 
\label{fig:entgen1}
\end{figure}

We simulate the entanglement generation rates and secure key rates for BBM92 in a scenario where entanglement is distributed between two end nodes via a repeater network containing $r$ repeater nodes. The repeater protocol begins after a synchronization message is sent by Alice to each node indicating the timestamp when all protocols should be started.  Each node then independently runs its own protocol to first generate entanglement across the elementary links. For each elementary link, the associated nodes each entangle a photonic qubit with a stationary qubit in their quantum memory, and then send the photonic qubit to a Bell state measurement station (BSM) located in the center of the elementary link. The photonic qubits are measured at the Bell state measurement station and the results transmitted to the adjacent nodes.

When elementary link entanglement has been established and the corresponding nodes have been informed,  the nodes then execute entanglement swapping to achieve end-to-end entanglement between Alice and Bob. Measurement results from entanglement swapping operations are collected and sent to Alice, who will perform a final correction on her qubit to complete the establishment of a shared and \textit{a priori} agreed-upon Bell state between Alice and Bob.

For entanglement generation protocols with deterministic swapping, there exists an exact analytical expression for the expected time to generate entanglement between Alice and Bob. The average time required to generate entanglement between the end nodes of a repeater network with $2^n$ elementary links and with an elementary link distance of $L_0$ is  \cite{PhysRevA.83.012323}
\begin{align}
\langle T \rangle = \frac{L_0}{c} \sum_{j=1}^{2^n} \binom{2^n}{j} \frac{(-1)^{j+1}}{1 - (1 - P_0)^j}
\end{align}
where $c$ is the speed of light and $P_0$ is the probability of establishing entanglement across an elementary link. $P_0 =  \frac{1}{2}\eta_\text{link}$, where $\eta_\text{link}$ is the transmission efficiency from one node to the Bell state measurement station (BSM) in the center of the elementary link, to the other node. For simplicity, we assume the photons travel in a vacuum, but to consider another material, the speed of light can be divided by the index of refraction of the material. Note the distance between a node and a BSM is $L_0/2$, and the total end-to-end length is $L = L_0 2^n$. There are two end nodes and $r = L/L_0 - 1$ repeater nodes. Additionally, for the elementary link entanglement generation, the Bell state measurement efficiency is one half, the efficiency of a standard linear optical Bell state measurement. The minimum classical communication time  to determine whether the entanglement generation attempt was successful after each attempt is also incorporated.

\begin{figure}[tb]
\includegraphics[width=8.5cm]{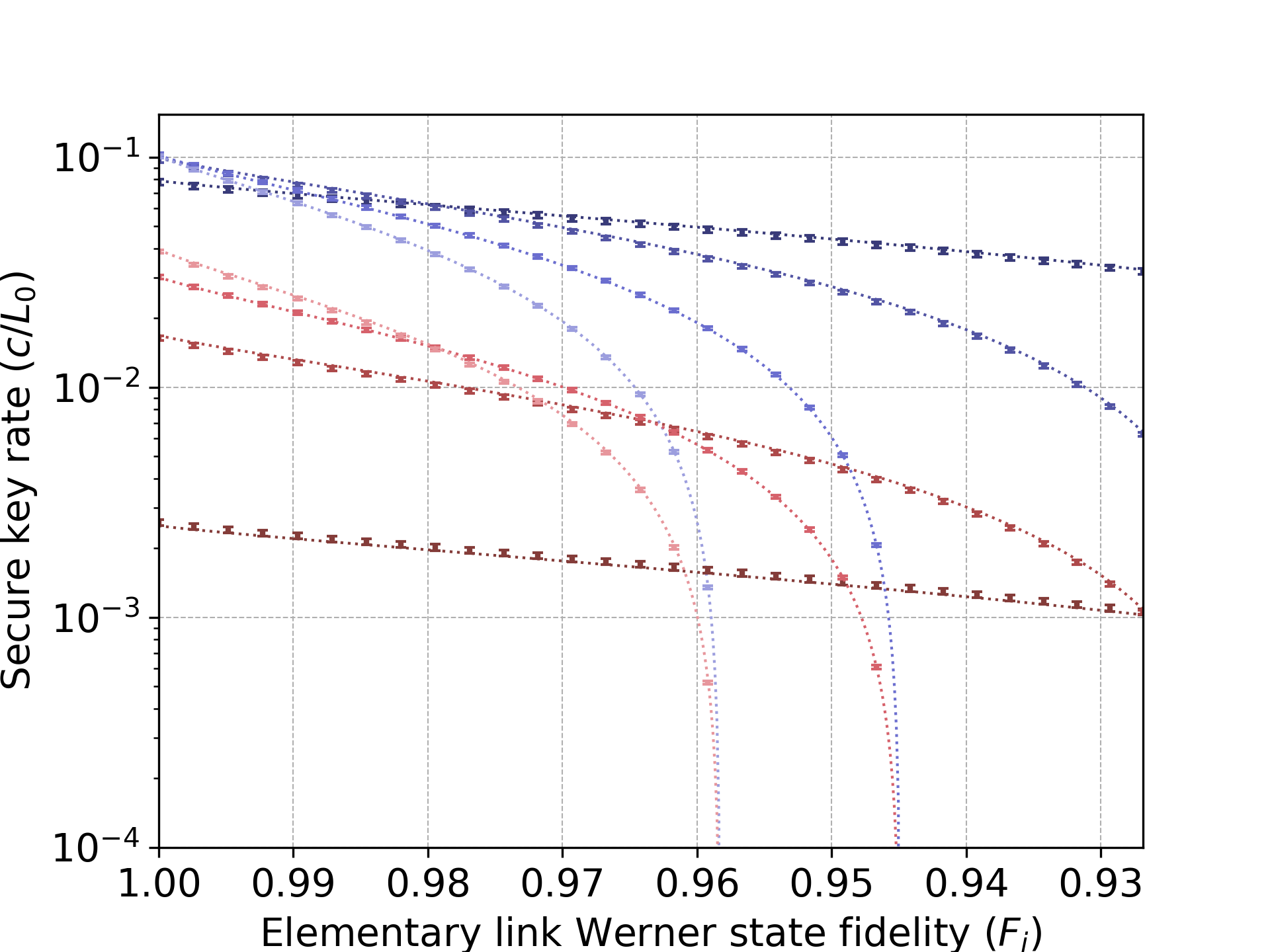}
\caption{Simulated (1000 shots) secure key rates  for a quantum repeater network with different numbers of repeater nodes, for different end-to-end losses, and for different elementary link Werner state fidelities $F_i$. The dotted lines show theoretical calculations while the error bars show AQNSim simulations. The red lines are repeater networks with 20 dB end-to-end loss, while blue lines are repeater networks with 5 dB end-to-end loss, with darker to lighter lines representing increasing numbers of repeater nodes [0, 1, 2, 3].  } 
\label{fig:entgen2}
\end{figure}

To validate our simulations, we plot the  average entanglement generation rate against elementary link loss in the case of $r = 1$ for theory and simulation (AQNSim). We assume deterministic swapping. The AQNSim results concur with theory given a Bell pair source with perfect fidelity (see Figure \ref{fig:entgen1} b).

Applying our simulator to longer repeater networks and to the specific use case of BBM92, we simulate entanglement generation rates and predict secret key rates for  a repeater network given deterministic swapping but imperfect elementary entanglement link fidelities  (see Figure \ref{fig:entgen2}). The theory curves use the expression for the final fidelity $F$ of an end-to-end entangled pair in a repeater network generated from elementary links with entangled states which are Werner states with fidelity $F_i$ \cite{Briegel1998}
\begin{align}
F = \frac{1}{4} + \frac{3}{4} \left( \frac{4 F_i - 1}{3} \right)^{r +1}.
\end{align}
The theory curves in Figure \ref{fig:entgen2} match the AQNSim simulations, verifying AQNSim simulation against  analytical theory for repeater networks with deterministic swapping.


Figure  \ref{fig:entgen2} provides insight into tradeoffs inherent in repeater networks. As the elementary link fidelity degrades, the secure key rate decreases, and it decreases at different rates for different repeater node numbers $r$. The optimum number of repeater nodes $r$ should be chosen based on the total end-to-end loss and the elementary link Werner state fidelity. 

Although the simulations plotted in Figure  \ref{fig:entgen2} are kept general and omit implementation-specific details including detector jitter, it is straightforward to scale up the model in complexity and detail. In future work, one can incorporate quantum gate errors and delays, user-specified classical signal delays, other noise models beyond depolarizing noise including time-varying noise models, finite quantum memory times, entanglement purification protocols, and other realistic details and extensions which may not be readily addressed purely using analytical theory.

There is no exact expression for entanglement generation rates in repeater networks given probabilistic swapping \cite{PhysRevA.100.032322}. Future work can also extend the deterministic swapping entanglement generation simulations discussed in this section to address entanglement generation in repeater networks given probabilistic swapping.

\section{Conclusions}

We have employed a discrete  event simulation framework for quantum networks  implemented using the Python package AQNSim, and shown it can accurately simulate key rates and QBERs for BBM92 key exchange. We have experimentally conducted BBM92 key exchange, reproduced the experimental setup with AQNSim, and shown our simulation results and experimental results align with a smaller mean squared error than experiment and analytical theory. Additionally, we have used AQNSim to solve for quantum network metrics in more complex repeater scenarios where experimental implementations do not yet exist, verifying our simulations against simple but exact theoretical models. 

We use AQNSim as a reference implementation of a discrete event simulation framework. Comparable quantum network Python simulation libraries exist and may be used for alternative implementations of the simulations presented in our work, provided they support the required features (see Appendix D).

In future work, discrete event simulation software with the requisite features and capabilities can be employed to simulate repeater protocols with additional complexity beyond that which is addressed in this work and with increasing levels of realism, including protocol-specific classical message delays, quantum memories with finite storage times, chromatic dispersion and detector jitter, detector dark counts, and probabilistic swapping.

Accurate, scalable, and modular quantum network simulators are critical tools for designing quantum networks and predicting protocol performance -- especially as network complexity scales to the point where theoretical expressions are lacking or are otherwise impractical to derive. 
As quantum technology continues to mature,  discrete  event quantum network simulators will accelerate real-world quantum secure communication deployments and advance related quantum applications including quantum sensing and distributed quantum computing.

\section*{Code and Data Availability}
All experimental data and code  other than the source code for the proprietary AQNSim Python package are openly available \cite{ourdata}. 

\section*{Acknowledgments}
The authors thank Ralf Riedinger, Marty Fitzgerald, and Oliver Stone for helpful discussions. We thank Michelle Fernandez for her contributions to the AQNSim Python package.

\appendix
\section{Equipment and synchronization}
\label{sec:appendixsync}

The entangled photon source was purchased from OZ Optics (EPS-1000-3U3U-810-5/125). The detectors used were single photon detection modules (SPDMH3) from Thorlabs and the time tagger was a Time Tagger Ultra by Swabian Instruments.

Temporal synchronization between detector pairs is necessary before timestamps and bases can be exchanged in a BBM92 protocol. Temporal synchronization is performed during setup calibration by finding the displacement of $g^{(2)}$ peaks for different detector pairings between Alice and Bob, and solving for the delays.

\section{Measurement correlation matrix}
\label{sec:appendixcorrelation}

\begin{figure}[tb]
\includegraphics[width=8.0cm]{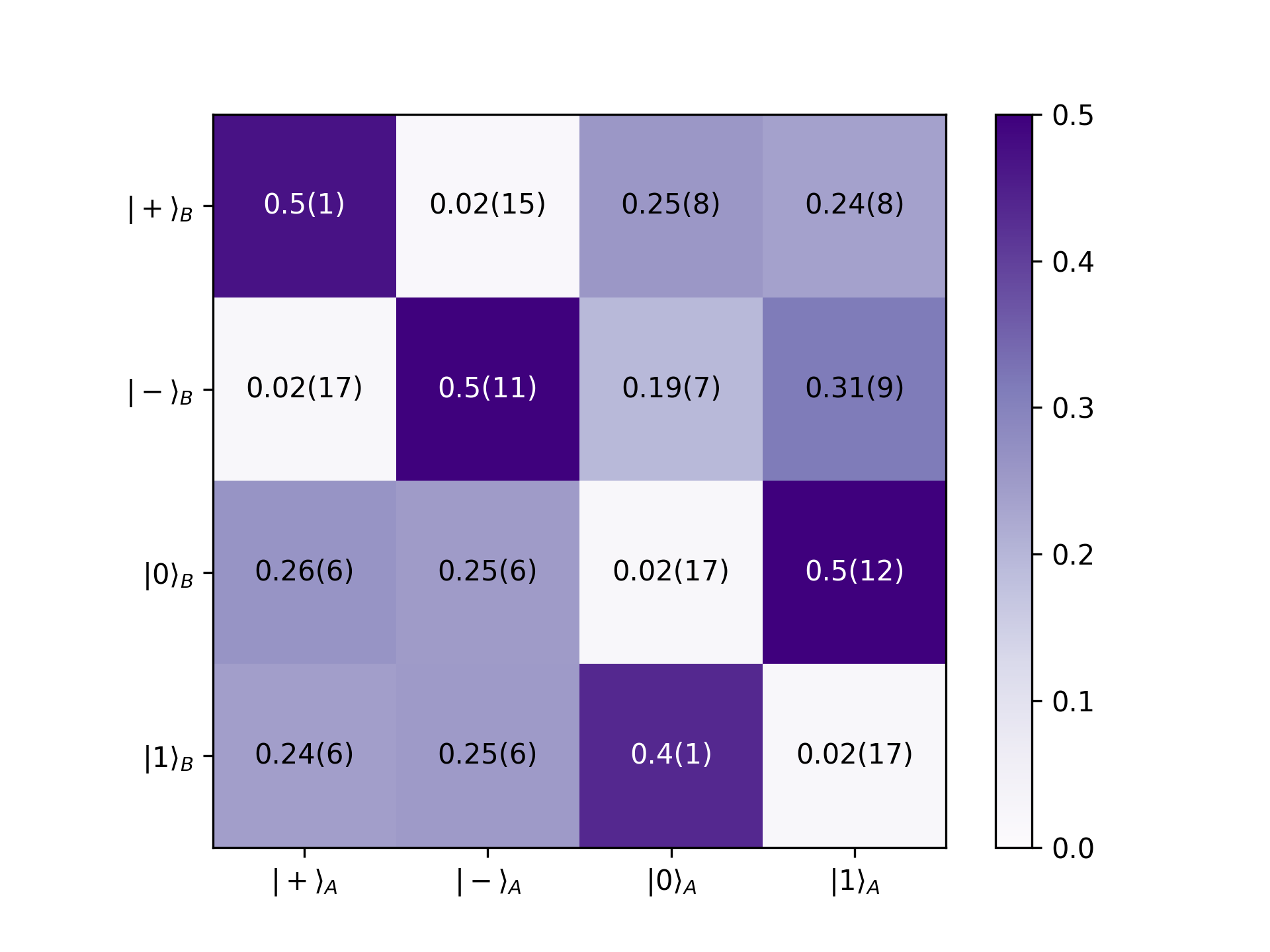}
\caption{ Correlation matrix calculated from experimental BBM92 data displaying the probabilities of Alice and Bob measuring a given outcome in a given basis. The coincidence window is 2 nanoseconds and other relevant experimental parameters are provided in the main text.} 
\label{fig:corrmatrix}
\end{figure}

The security of BBM92 is dependent upon the orthogonality of the two mutually unbiased measurement bases that Alice and Bob use to perform measurements. We quantify the bases orthogonality in our BBM92 experiment by plotting the probabilities of correlation for each measurement outcome and basis for Alice and Bob (see Figure \ref{fig:corrmatrix}). The  measurement correlation matrix for the Bell state $\ket{\Psi^+}$ has perfect correlations in the XX and ZZ  subspaces, and no correlations in the XZ and ZX  subspaces.

\section{Parameter estimation from experimental data}
\label{sec:appendixparameter}

\begin{figure}[tb]
\includegraphics[width=8.0cm]{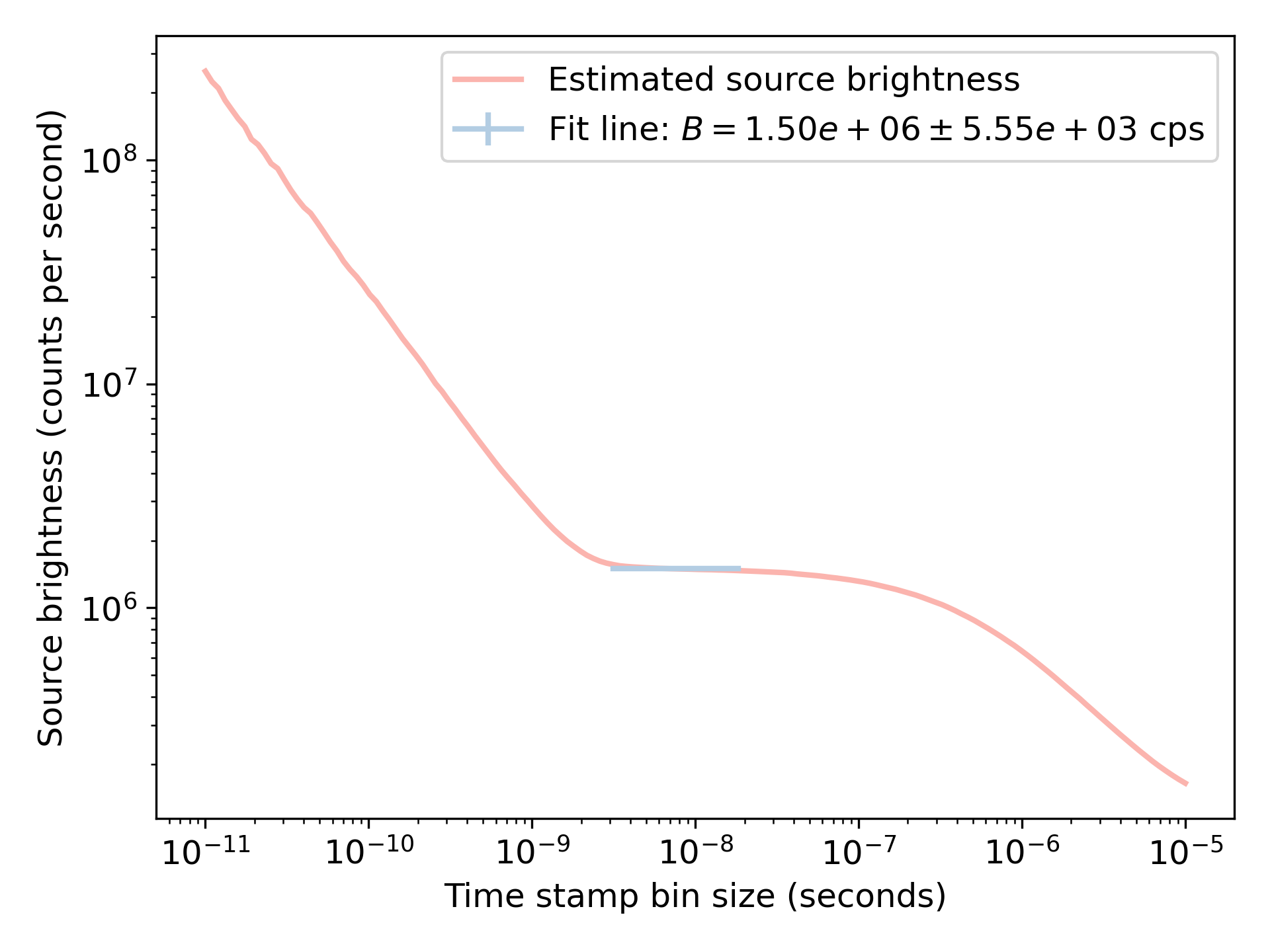}
\caption{ Estimate of internal source brightness (following equation \ref{eq:brightness}) plotted against coincidence window size. A linear fit for coincidence windows larger than the detection resolution provides an estimate for the internal source brightness in coincidences per second.  } 
\label{fig:sourcebrightness}
\end{figure}

The internal source brightness (the source brightness prior to coupling losses) must be estimated in order to simulate key rates and QBERs both with a theoretical model and using AQNSim. An estimate for the source brightness in a regime where detection resolution is several times less than the size of the coincidence window and where accidental coincidences contribute negligibly to measured coincidence counts may be found from the coincidence to accidental ratio, \textit{CAR}. The source brightness $B$ relates to the \textit{CAR} following the expression

\begin{align}
\label{eq:brightness}
B &\approx \frac{1}{\textit{CAR} \; t_{c}}  \\&\approx \frac{(S_A-D_A) (S_B-D_B)}{R_{\text{coinc}}(t_{c})}
\end{align}
Here $R_{\text{coinc}}(t_{c})$ is written as a function of $t_{c}$ to emphasize that it is a function of the coincidence window used for determining correlated counts.

For our BBM92 data, we plot the estimated internal source brightness against coincidence window size (see Figure \ref{fig:sourcebrightness}). The brightness is estimated using experimentally measured coincidence count rates at varying coincidence window sizes and the singles count rates at Alice and Bob (following equation \ref{eq:brightness}). A linear fit is performed in the region of this plot where accidental coincidences are unlikely compared to true coincidences, and where the coincidence window is greater than the detection jitter to estimate $B$.

Theoretical and simulated models for BBM92 key rates and QBERs also require an estimate of detection resolution.  Detection resolution for each detector pair during setup calibration is estimated as the average FWHM of each $g^{(2)}$ corresponding to detector pairings between Alice and Bob for matching bases and corresponding bits. In the case of more widely varying $g^{(2)}$ FWHMs between detector pairs, the  simulation model can be easily modified to take these varying peak widths into account.

\begin{table*}[htb]
\centering
\caption{Comparison of different discrete event quantum network Python libraries by the features necessary for the simulations presented in the main text of this work. \checkmark indicates support for feature as found in the simulator's source code or documentation; X indicates no support for feature.} 
\begin{tabular}{l|l|l|l|l|}
\cline{2-5}
                                                                                                                      & \textbf{AQNSim} & \textbf{SeQUeNCe} & \textbf{QuISP} & \textbf{NetSquid} \\ \hline
\multicolumn{1}{|l|}{\begin{tabular}[c]{@{}l@{}}Continuous-wave polarization-\\ entangled photon source\end{tabular}} & \checkmark             & X                & X             & \checkmark              \\ \hline
\multicolumn{1}{|l|}{Beam splitter component}                                                                         & \checkmark             & \checkmark               & X             & X              \\ \hline
\multicolumn{1}{|l|}{Waveplate component}                                                                             & \checkmark             & X                & X             & X              \\ \hline
\multicolumn{1}{|l|}{Link photon loss}                                                                                & \checkmark             & \checkmark               & \checkmark            & \checkmark               \\ \hline
\multicolumn{1}{|l|}{Detector dead time}                                                                              & \checkmark             & \checkmark               & \checkmark            & \checkmark              \\ \hline
\multicolumn{1}{|l|}{Detection timing jitter}                                                                         & \checkmark             & \checkmark               & X             & \checkmark              \\ \hline
\multicolumn{1}{|l|}{Detector dark counts}                                                                            & \checkmark             & \checkmark               & \checkmark            & \checkmark               \\ \hline
\multicolumn{1}{|l|}{\begin{tabular}[c]{@{}l@{}}Parallelization option \\ for many simulation runs\end{tabular}}      & \checkmark             & \checkmark               & X             & \checkmark               \\ \hline
\end{tabular}
\label{tab:supptable}
\end{table*}

\begin{figure}[tb]
\includegraphics[width=8.0cm]{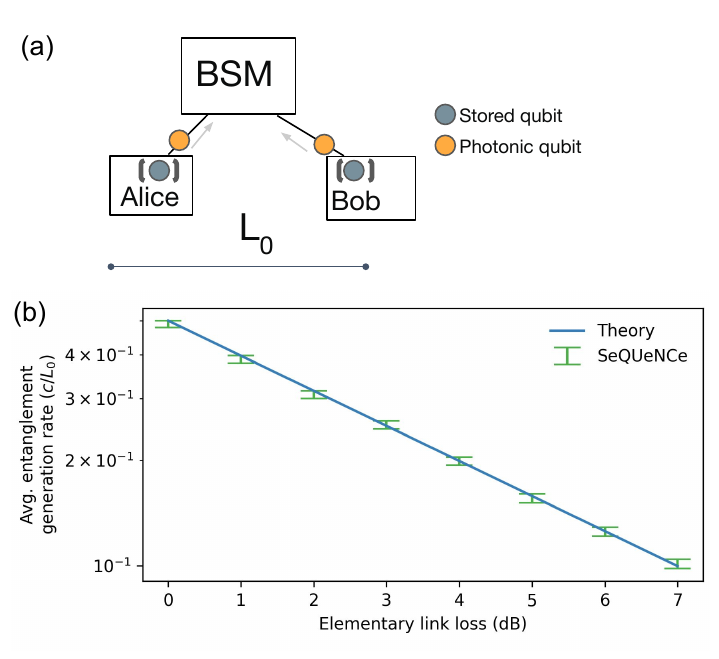}
\caption{(a) Entanglement creation between two end nodes (Alice and Bob), with no repeater nodes ($r=0)$, and a single Bell state measurement station (BSM).  (b) The average entanglement generation rate (in units of $c/L_0$) for different elementary link loss values calculated from simulation (using SeQUenCE with 1000 shots) and from theory.}
\label{fig:sequence}
\end{figure}

The probability of error due to optical imperfections, $p_0$, is estimated as the experimentally calculated QBER using a coincidence window of 1 nanosecond, such that the ratio of accidental coincidences to true pair coincidences is negligible. This QBER is effectively a measurement of the entangled photon source and optical setup visibility, which is a visibility that incorporates quantum state errors due to any defects in the measurement optics and integrated source and due to any imperfections in the polarization drift compensation. This estimate of the source visibility is corroborated by independently taken visibility measurements of the entire optical setup, including the entangled photon source and the measurement optics.

\section{Additional quantum network simulation platforms}
\label{sec:appendixothersimulators}
We use the discrete event quantum network simulation Python library AQNSim for the simulations presented in our work. There are many other existing discrete event quantum network simulation Python libraries, many with comparable features and functionalities [25-28]. The simulation framework presented in this paper may, in principle, be implemented with another discrete event quantum network simulator with the necessary features. Table \ref{tab:supptable} lists the features needed to reproduce all simulations in this work while noting their presence in quantum network simulators AQNSim, SeQUeNCe [25], QuISP [26], and NetSquid [27]. As the table shows, many of the features necessary for our work are common across the other listed simulators. However, exactly replicating all of the simulations presented in this work would require custom implementation of new components or features if using a discrete event quantum network simulator other than AQNSim.

SeQUeNCe is a commonly used quantum network discrete event simulator which is open-source with several basic built-in protocols and a sizable component library. To demonstrate another example of validation for another quantum network simulation platform, we constructed a basic entanglement distribution simulation using SeQUeNCe. For the simulation, Alice and Bob are each located at a distance $L_0/2$ from a central Bell state measurement station and each generate shared entanglement across their quantum memories using a simple entanglement generation protocol (see Figure \ref{fig:sequence}a). The average entanglement generation rates are plotted in Figure \ref{fig:sequence}b against the total link loss (total loss from Alice to the BSM to Bob). We additionally incorporate the minimum classical communication time to determine entanglement generation attempt success in the entanglement generation rates. We see agreement of simulation with theory using the same analytical model used for the simulations shown in Figure 5b in our main text, here with repeater node number $r=0$.

We note that the SeQUeNCe implementation for a Bell state measurement between two photonic qubits entangled with quantum memories, SingleAtomBSM, as of the writing of this paper, applies the channel loss only for one beam splitter input arm, either Alice to BSM or Bob to BSM, when only a single photon is measured.  In this work, we model entanglement generation protocols in which a photon is transmitted along each channel towards the beam splitter and each photon independently experiences channel loss. To this end, we implemented a modified version of the SingleAtomBSM class to apply channel loss independently for both arms, Alice to BSM and Bob to BSM.

\medskip
\bibliographystyle{unsrt}
\bibliography{sources}

\end{document}